\documentclass[letterpaper, 10pt, conference]{ieeeconf}
\IEEEoverridecommandlockouts  
\usepackage{array}
\usepackage{cite}
\usepackage{amsmath,amssymb,amsfonts}
\usepackage{graphicx}
\usepackage{textcomp}
\usepackage{xcolor}
\usepackage{booktabs} 
\usepackage{float}
\usepackage{times}
\usepackage{soul}
\usepackage{url}
\usepackage[hidelinks]{hyperref}
\usepackage[utf8]{inputenc}
\usepackage[small]{caption}

\usepackage{booktabs}
\usepackage{algorithm}
\usepackage[switch]{lineno}
\usepackage{makecell} 
\usepackage{algpseudocode}
\DeclareMathOperator*{\argmin}{arg\,min}
\def\BibTeX{{\rm B\kern-.05em{\sc i\kern-.025em b}\kern-.08em
    T\kern-.1667em\lower.7ex\hbox{E}\kern-.125emX}}
\begin{document}

\title{A Game-Theoretic Framework for Intelligent EV Charging Network Optimisation in Smart Cities}

\title{\LARGE \bf
A Game-Theoretic Framework for Intelligent EV Charging Network Optimisation in Smart Cities
}

\author{
Niloofar Aminikalibar\textsuperscript{*}, Farzaneh Farhadi\textsuperscript{*}, and Maria Chli\textsuperscript{*}
\thanks{
\textsuperscript{*} Niloofar Aminikalibar, Farzaneh Farhadi, and Maria Chli are with the School of Computer Science and Digital Technologies, Aston University, Birmingham, UK. Emails: {\tt\small namin21@aston.ac.uk, f.farhadi@aston.ac.uk, m.chli@aston.ac.uk}}%
}
\maketitle
\begin{abstract}
The transition to Electric Vehicles (EVs) demands intelligent, congestion-aware infrastructure planning to balance user convenience, economic viability, and traffic efficiency. We present a joint optimisation framework for EV Charging Station (CS) placement and pricing, explicitly capturing strategic driver behaviour through coupled non-atomic congestion games over road networks and charging facilities. From a Public Authority (PA) perspective, the model minimises social cost, travel times, queuing delays and charging expenses, while ensuring infrastructure profitability.
To solve the resulting Mixed-Integer Nonlinear Programme, we propose a scalable two-level approximation method, Joint Placement and Pricing Optimisation under Driver Equilibrium  (JPPO-DE), combining driver behaviour decomposition with integer relaxation. Experiments on the benchmark Sioux Falls Transportation Network (TN) demonstrate that our method consistently outperforms single-parameter baselines, effectively adapting to varying budgets, EV penetration levels, and station capacities. It achieves performance improvements of at least 16\% over state-of-the-art approaches. A generalisation procedure further extends scalability to larger networks. By accurately modelling traffic equilibria and enabling adaptive, efficient infrastructure design, our framework advances key intelligent transportation system goals for sustainable urban mobility.

\end{abstract}


\section{Introduction}

The transition to Electric Vehicles (EV) is essential to global decarbonisation strategies \cite{bibra2023global}, with policies such as the UK’s planned phase-out of new petrol and diesel cars by 2030 \cite{news1}. However, widespread EV adoption depends not only on vehicle availability but also on the intelligent deployment of supporting infrastructure. In particular, the planning of fast EV charging networks must address traffic-aware siting, congestion mitigation, and economically sustainable operation.

A critical challenge is the “chicken-and-egg” problem: without sufficient demand, investment in charging infrastructure is economically risky; yet, without adequate infrastructure, EV uptake is limited \cite{RN58, shi2021comprehensive}. Moreover, poorly located charging stations can exacerbate road congestion, as drivers detour or queue in already busy areas \cite{9922278}. 

This paper addresses the design of EV charging networks that jointly optimise station placement and pricing, while explicitly modelling the strategic behaviour of drivers. EV users select routes and charging locations to minimise personal travel time, queuing delays, and financial cost. These decentralised choices interact to form user equilibrium across both roads and charging facilities, significantly affecting system-wide performance.

Our game-theoretic bi-level optimisation framework minimises social cost (aggregate travel time, delay, and cost) while ensuring infrastructure profitability. Specifically, it:
(1) supports Public Authorities (PA) and urban planners in strategically placing chargers while accounting for congestion at both road and station levels;
(2) offers pricing strategies that balance demand, congestion management, and economic sustainability; and
(3) provides actionable insights into the spatial distribution of EV and non-EV traffic to inform policy and intervention design.

To address scalability, we introduce a two-level approximation algorithm, Joint Placement and Pricing Optimisation under Driver Equilibrium (JPPO-DE), that decomposes traffic flow analysis and relaxes placement variables, enabling tractable solutions on city-scale networks. We evaluate our approach on the benchmark Sioux Falls transport network~\cite{leblanc1975}, demonstrating reductions in social cost compared to single-parameter baselines and adaptability under varying budgets, EV penetration, and infrastructure disruptions. These results offer practical insights for sustainable urban mobility planning.

\section{Literature Review} \label{sec:literature}

Recent advances in the Intelligent Transportation Systems (ITS) community \cite{RN57, 9922278, 8317680, 9565089, 10421957, RN62, RN153, zhao2021dynamic, RN140} have explored optimisation methods for EV charging infrastructure, addressing placement, pricing, or both. However, most studies consider these elements in isolation, missing the interaction between infrastructure siting, user behaviour and congestion effects.

\textbf{Placement optimisation:}
Numerous studies aim to improve accessibility or reduce travel distances through optimal charger location. Reviews such as \cite{RN57} categorise these works by objective and method. Examples include \cite{9922278, 8317680}, minimising detours and waiting times; \cite{9565089}, combining placement with fleet routing; and \cite{10421957}, which integrates travel time and energy consumption, highlighting growing attention to user experience.

\textbf{Pricing optimisation:} Other studies explore dynamic pricing for profit maximisation or demand balancing, using reinforcement learning to adapt to usage patterns \cite{RN62,RN153}, reduce service rejections \cite{zhao2021dynamic}, or enhance user satisfaction \cite{RN140}.

\textbf{Joint placement and pricing:} A limited set of works addresses both siting and pricing. For example, \cite{RN176} uses a Stackelberg game, and \cite{RN155} employs sequential optimisation, but both neglect user interaction and equilibrium dynamics. These models often assume drivers act in isolation or passively follow system decisions.

\textbf{Strategic driver behaviour:} In practice, EV users behave selfishly, routing and charging to minimise personal cost. This leads to user equilibrium, which differs from centralised optima  \cite{wang2015analysis, youn2008price}. Congestion games provide a natural framework to model such interdependence \cite{RN178}, and have been applied in various ITS contexts  \cite{10591217, RN167, xiong2016optimal, bakhshayesh, zavvos2021}. However, most focus on either road \cite{10591217, xiong2016optimal} or station congestion \cite{RN167}, not both.

\textbf{Our contribution:} Building on this foundation, we propose the first ITS-oriented framework that jointly optimises EV charger placement and pricing under a \emph{coupled congestion game}, capturing network-wide interactions between strategic driver decisions, road traffic, and station queuing. Our approach supports \emph{realistic deployment} by integrating multi-agent behaviour, traffic dynamics, and economic constraints into a single optimisation model. This enables adaptive planning, aligns economic and social objectives, and facilitates resilient, scalable infrastructure strategies for smart urban transport.

\section{System Modelling and Intelligent Optimisation Framework} \label{sec:section2}
\begin{table}[ht]
\centering
\caption{Notation Guide}
\label{table:symbols}
\begin{tabular}{|>{\raggedright\arraybackslash}m{0.026\textwidth}|m{0.164\textwidth}|>{\raggedright}m{0.026\textwidth}|m{0.164\textwidth}|}
\hline
\textbf{Term} & \textbf{Description} & \textbf{Term} & \textbf{Description} \\
\hline
$\omega$ & O-D pair: $\omega \in W$ & $i$ & Node identifier \\
\hline
$r_\omega$ & A route for $\omega$ & $p_\omega$ & An extended path for $\omega$ \\
\hline
$R_\omega$ & Set of routes for $\omega$ & $P_\omega$ & Set of extended paths for $\omega$ \\
\hline
$q_{\omega,p}$ & proportion of EV drivers with $\omega$ choosing extended path $p$ & $\boldsymbol{q_{\omega}}$ & Strategy over all extended paths for O-D $\omega$ \\
\hline
$q_{\omega ,r}^0$ & proportion of NCDs with $\omega$ choosing route $r$  & $\boldsymbol{q_{\omega}^0}$ & Strategy over all routes for O-D $\omega$ \\
\hline
$x_i$ & Number of chargers at node $i$ & $y_i$ & Charging price at node $i$ \\
\hline
$e_i$ & Wholesale electricity price at node $i$ & $T_i$ & Rental/maintenance cost at node $i$ during peak hours \\
\hline
$\gamma_\omega$ & Flow of en-route-charging EVs for $\omega$ & $\gamma_\omega^0$ & Flow of Non-Charging Drivers for $\omega$ \\
\hline
$d_l$ & Distance of link $l$ & $c_l$ & Capacity of link $l$ \\
\hline
$\lambda$ & Monetary value of time/h& $\mu$ & Service rate per charger/h \\
\hline
$B$ & Budget required per charger& $\pi$ & Profit Coefficient \\
\hline
$\alpha$ & Penetration rate & $\boldsymbol{Q}$ & Strategy profile of all drivers across all O-Ds \\
\hline
\end{tabular}
\end{table}

We propose a computational framework combining \emph{game-theoretic behaviour modelling} and \emph{bi-level optimisation} to support intelligent EV infrastructure planning. Our hybrid approach integrates \emph{multi-agent decision-making}, \emph{mixed-integer programming}, and \emph{network congestion dynamics}, offering a scalable foundation for smart city traffic management. This framework is modular and can be extended to support heterogeneous energy demands, partial route charging, or competition between Charging Stations (CS) operators.

This section first describes the Transportation Network (TN) and agent types. It then outlines the behavioural drivers behind routing and charging choices. Finally, we formalise the PA's objective and the resulting game structure.

\subsection{Transportation Network Representation}\label{sec:Charging Network Architecture}

The TN is modelled as a connected directed graph \( G = (N, A) \), where \(N\) denotes the set of nodes (urban locations) and \(A\) denotes the set of directed links (roads). Each link \( l \in A \) has an associated distance \( d_l \) and capacity \( c_l \), where higher capacity corresponds to greater vehicle throughput and reduced congestion levels.

Vehicles travel across the network between origin-destination (O-D) pairs \( \omega \). For each O-D pair, \( R_\omega \) denotes the set of feasible routes connecting the origin to the destination. To model EV charging decisions, we define the set of \emph{extended paths} \( P_\omega \), where each extended path \( p \in P_\omega \) specifies both a driving route and a selected charging location along that route. For example, if a route \( r \in R_\omega \) is represented by the sequence of nodes $r=(1,2,3,4)$, an extended path would be a pair \(p=(r, i)\), where \( i \in \{1,2,3,4\} \) indicates the node where the EV driver chooses to charge.

We introduce the function \( s(p) \) that maps each extended path \( p \) to the corresponding node \( i \) where the driver chooses to charge (e.g., \( s(p) = i \) in the example above). This structure allows the model to simultaneously capture congestion effects on roads and at charging stations, enabling a more integrated optimisation of traffic dynamics and infrastructure usage.

\subsection{Agents and Decision Variables}

We model three interacting agents: the PA, Non-Charging Drivers (NCDs), and en-route charging EV drivers.

The PA is responsible for infrastructure decisions, determining the number of chargers \( x_i \) to install at each node \( i \in N \) and setting the corresponding charging price \( y_i \geq 0 \). Each location \( i \) is associated with a wholesale electricity price \( e_i \) (per EV charge) and fixed maintenance and rental fees \( T_i \) (per CS) over the modelling period. PA decisions must respect a global budget \( B \), cover location-specific costs, and ensure profitability through appropriate pricing.

NCDs represent background traffic, including non-EV users and EVs that do not require en-route charging. For each O-D pair \( \omega \), their traffic is distributed over the available routes \( R_\omega \) according to a flow distribution vector \( \mathbf{q}_\omega^0 = (q_{\omega,r}^0)_{\forall r} \), where \( q_{\omega,r}^0 \) denotes the proportion of NCDs selecting route \( r \). This formulation captures aggregate flow behaviour rather than individual decision-making.\footnote{
Superscript  \( 0 \)
indicates NCD-specific parameters in the paper.}

EV drivers with charging needs select extended paths \( p \in P_\omega \) that jointly determine their routing and charging decisions. Their strategy is represented by \( \mathbf{q}_\omega = (q_{\omega,p})_{\forall p}\), where $q_{\omega, p}$ is the proportion of EV drivers selecting extended path $p$.

\subsection{Driver Cost Components}\label{sec: behaviour factors}

Both EV drivers and NCDs aim to minimise their individual travel costs when traversing the network. For EV drivers, the total cost includes three main components: travel time, queuing time at CSs, and financial cost for charging, while NCDs are affected only by travel time. Importantly, drivers' decisions are interdependent: each individual's choice affects overall congestion levels, which in turn influence the costs faced by others. This strategic interdependence across the network necessitates an equilibrium-based modelling approach to align individual behaviours with system-wide outcomes.

\subsubsection{ Travel Time}
The travel time incurred along a link \( l \) depends on the level of congestion and is given by:
\begin{equation}
f_l = \frac{d_l (\gamma_l^0 + \gamma_l)}{c_l},
\end{equation}
where \( d_l \) is the distance of link \( l \), \( c_l \) is its capacity, and 
\begin{equation}
\gamma^0_l=\sum_{\omega, r : r \in R_{\omega}, l \in r} \gamma^0_{\omega} q^0_{\omega, r}, \qquad\gamma_l=\sum_{\omega, p : p \in P_{\omega}, l \in p} \gamma_{\omega} q_{\omega, p},
\end{equation}represent the expected NCD and EV drivers' traffic flows, respectively.\footnote{The linear travel time function is justified by the Conical Delay Function, which approximates travel delay under moderate congestion \cite{spiess1990conical}.} 
The total travel time for a driver following a route \( r \in R_\omega \) or an extended path \( p \in P_\omega \) is obtained by summing the travel times across all traversed links:
\begin{equation}
F_{\omega,r} = \sum_{l \in r} f_l, \quad F_{\omega,p} = \sum_{l \in p} f_l.
\end{equation}

\subsubsection{Queuing Time at Charging Stations}

EV drivers also experience queuing delays at CSs, which depend on station-specific congestion. The expected waiting time at the CS associated with path \( p \) is modelled as:
\begin{equation}
G_p = \frac{\sum_{\omega, p': s(p') = s(p)} \gamma_{\omega} q_{\omega,p'}}{\mu x_{s(p)}} \quad \forall p, p' \in P_\omega,
\end{equation}
where the numerator represents the expected arrival rate of EVs choosing to charge at station \( s(p) \), \( \mu \) is the service rate of a single charger per hour reflecting number of EVs fully charged by a charger in an hour, and \( x_{s(p)} \) is the number of chargers installed at that station\footnote{The expected queueing time at a CS is approximated using an aggregated queuing delay model based on service rate and arrival intensity. Although this does not explicitly model individual queues (e.g., M/M/c), it captures congestion effects at the network level and aligns with non-atomic traffic flow assumptions.}. As traffic volume increases or available chargers decrease, expected queuing times correspondingly rise.

\subsubsection{Financial Charging Cost}\label{sec:financial}

In addition to time-related costs, EV drivers incur a financial cost for charging their vehicles. This cost depends solely on the pricing decision at the selected CS and is given by:
\begin{equation}
Y_p = y_{s(p)},
\end{equation}
where \( y_{s(p)} \) is the price set at the station corresponding to the charging decision along extended path \( p \). 

For simplicity, we assume all EVs have identical energy demand, resulting in uniform charging durations and prices across users at a given station.


\subsubsection{Total Driver Cost} \label{sec:cost-functions}

The overall cost for an EV driver choosing extended path \( p \) combines all components and is given by:
\begin{equation}
C_{\omega,p}(\mathbf{Q}) = \lambda (F_{\omega,p} + G_p) + Y_p,
\end{equation}
where \( \lambda \) is a parameter converting time units into monetary units for aggregation, and \( \mathbf{Q} = (\mathbf{q}_{\omega}, \mathbf{q}_{\omega}^0)_{\forall \omega} \) denotes the full strategy profile of EV drivers and NCDs across all O-D pairs. Hence, the expected cost for EV drivers with O-D pair \( \omega \) is:
\begin{equation}\label{eq-EVcost}
C_{\omega}(\mathbf{Q}) = \sum_{p \in P_\omega} q_{\omega,p} C_{\omega,p}(\mathbf{Q}).
\end{equation}

Similarly, the travel cost incurred by an NCD travelling on a route \( r \in R_\omega \) is:
\begin{equation}
C_{\omega,r}^0(\mathbf{Q}) = \lambda F_{\omega,r}.
\end{equation}
Accordingly, the expected cost for NCDs with O-D pair \( \omega \) is:
 \begin{equation}\label{eq-NonEVcost}
C_{\omega}^0(\mathbf{Q}) = \sum_{r \in R_\omega} q_{\omega,r}^0 C_{\omega,r}^0(\mathbf{Q}).
\end{equation}

\subsection{Public Authority Objective: Bi-Level Optimisation} \label{sec:CSO-objective}
We assume that PA owns and manages all charging CSs across the network. This centralised ownership enables the PA to coordinate infrastructure siting and pricing decisions to minimise the total social cost, comprising travel delays, queuing times, and charging expenses, while ensuring operational profitability. The PA must operate within a fixed budget and maintain a minimum profit margin at each station, balancing economic viability with public service objectives.


While drivers act selfishly to minimise their own costs, the PA aims to minimise the overall social cost:
 \begin{equation}
\Theta(\mathbf{Q}) = \sum_{\omega} \left( \gamma_\omega C_\omega(\mathbf{Q}) + \gamma_\omega^0 C_\omega^0(\mathbf{Q}) \right).
\end{equation}

The PA's optimisation problem is formulated as a \emph{bi-level program}, where some of the constraints themselves involve solving optimisation problems. Specifically, the PA chooses the placement \( x_i \) and pricing \( y_i \) decisions at the upper level, anticipating that drivers will react optimally by selecting routes and CSs to minimise their own costs at the lower level.

\begin{equation}\label{CSO Problem}
\begin{aligned}
&\min_{x,y,\mathbf{Q}} \Theta(\mathbf{Q}) \hspace{3cm} \text{\textbf{Bi-Level PA Problem}}\\
  s.t:  
        &\sum_{i \in N} x_{i}  \leq B, \\ 
      &\pi[\sum_{\omega, p: s(p)=i} \left(\gamma_{\omega} q_{\omega,p} e_{i}\right) + x_{i} T_{i}] \leq \sum_{\omega, p: s(p)=i} \gamma_{\omega} q_{\omega, p} y_{i},\forall i,\\
      &\mathbf{q}_\omega \in \argmin_{\mathbf{q}_\omega} C_\omega \mathbf{(Q)}, \quad \forall \omega, \hspace{0.3cm} \text{EV Driver Equilibrium}\\
      &\mathbf{q}_\omega^0 \in \argmin_{\mathbf{q}_\omega^0} C_{\omega}^0(\mathbf{Q}), \quad \forall \omega, \hspace{1cm} \text{NCD Equilibrium}\\
    & x_{i} \in \mathbb Z{\ge 0},\qquad y_i \geq 0  \quad \forall i,\\
    &0 \leq q_{\omega, p} \leq 1\quad \forall p,\omega,\qquad \sum_p q_{\omega,p}=1 \qquad \forall \omega, \\
    &0 \leq q_{\omega,r}^0\leq 1\quad \forall r,\omega, \qquad  \sum_r q^0_{\omega,r}=1 \qquad \forall \omega.
  \end{aligned}
\end{equation} 

The first set of constraints enforces the budget limit, restricting the total number of chargers that can be installed based on available funding. The second set ensures profitability at each CS: the total revenue from EV drivers must cover the electricity costs \( e_i \), maintenance and rental costs \( T_i \), while achieving at least a profit margin \( \pi > 1 \) (e.g., \( \pi = 1.2 \) guarantees a 20\% profit margin).

The Driver Equilibrium (DE) constraints, for both EV and NCD drivers, ensure that agents' behaviour aligns with minimising their respective cost functions~\eqref{eq-EVcost} and~\eqref{eq-NonEVcost}. Each placement and pricing configuration induces a competitive game among drivers, and the PA must anticipate these strategic interactions when optimising the network design. The DE constraints formalise best-response behaviour: each driver selects a routing and charging decision from an \(\argmin\) set that minimises their individual cost given the current network congestion. This ensures that drivers' routing and charging decisions form a Nash Equilibrium (NE), where no driver can unilaterally lower their cost by changing their route or charging location. The detailed game-theoretic structure underlying these interactions is presented in the next section.

The final set of constraints ensures feasibility: placement decisions \(x_i\) must be integer-valued, pricing decisions \(y_i\) must be non-negative, and traffic distributions must form valid probability vectors across routes and extended paths. 

Overall, the system integrates multi-agent decision-making, with drivers acting as decentralised agents seeking to minimise individual costs, and mixed-integer programming, where the PA optimises discrete placement and continuous pricing decisions. By explicitly incorporating real-world factors such as budget limits, profitability requirements, and strategic driver behaviour, the framework supports scalable and adaptive EV infrastructure planning. This hybrid modelling approach fulfils the intelligent optimisation goals set out at the beginning of this section and aligns with broader ITS objectives of building resilient, efficient, and user-centric urban mobility systems.

\section{Strategic Traffic Dynamics and Intelligent Equilibrium Modelling}

The bi-level optimisation problem introduced in Section~\ref{sec:CSO-objective} requires anticipating how EV and NCD drivers respond to placement and pricing decisions. To simplify the PA problem, this section analyses the underlying \emph{driver interaction game}, studies its equilibrium properties, and derives explicit conditions for DE. By embedding these equilibrium conditions into the PA problem, the original bi-level program can be reformulated into a single-level optimisation model that remains tractable for intelligent infrastructure planning.

\subsection{Modelling Driver Interactions as Congestion Games}

The interaction between drivers in the TN naturally leads to competition for shared resources, road links and CSs. To capture this strategic behaviour, we model the system as two interdependent congestion games: a \emph{road congestion game}, in which all drivers (both EV and NCD) participate, and a \emph{CS congestion game}, affecting only EV users who require en-route charging. EV drivers simultaneously impact congestion on both roads and CSs, intertwining the two games.

Each driver aims to minimise their individual travel cost, which depends on congestion levels experienced across both domains. Given the large number of drivers relative to the impact of any single agent, it is appropriate to adopt a \emph{non-atomic congestion game} framework~\cite{milchtaich1996}, where each player's influence on aggregate congestion is negligible. This assumption aligns with real-world traffic systems, enabling a tractable yet realistic characterisation of network dynamics.

In the following subsection, we formalise the equilibrium conditions of the intertwined congestion games and derive criteria for stable driver behaviour across the network.

\subsection{Equilibrium Conditions for Driver Behaviour}\label{sec:nash}

In non-atomic congestion games, a strategy distribution constitutes an equilibrium if no infinitesimal agent can unilaterally reduce their cost by changing their decision. As established in~\cite{roughgarden2002bad}, an action distribution constitutes an NE if, for each driver type (EV or NCD with a given O-D pair) all strategies (i.e. extended paths for EVs and routes for NCDs) used with nonzero probability yield the minimum possible cost.

The equilibrium conditions for EVs and NCDs are then:
\begin{equation}\label{eq:NE-EV}
q_{\omega,p} C_{\omega,p}(\mathbf{Q}) \leq q_{\omega,p} C_{\omega,p'}(\mathbf{Q}), \quad \forall \omega, \forall p, p' \in P_\omega,
\end{equation}
\begin{equation}\label{eq:NE-NCD}
q^{0}_{\omega,r} C^0_{\omega,r}(\mathbf{Q}) \leq q^{0}_{\omega,r} C^0_{\omega,r'}(\mathbf{Q}), \quad \forall \omega, \forall r, r' \in R_\omega.
\end{equation}
These conditions ensure stable traffic patterns under decentralised, self-interested driver behaviour. Substituting the DE constraints in the Bi-Level PA problem with the explicit equilibrium conditions~\eqref{eq:NE-EV} and~\eqref{eq:NE-NCD} reformulates the problem as a single-level optimisation, where driver distributions adapt intelligently to placement and pricing decisions.

This equilibrium modelling not only improves infrastructure design accuracy but also enhances network resilience. By anticipating decentralised driver responses to congestion and pricing, the framework enables adaptive traffic redistribution under disruptions or demand changes, supporting ITS objectives of scalable, robust, and user-responsive mobility systems.

The next section presents the solution approach for the single-level PA problem, detailing the computation of optimal charger placement, pricing, and network traffic flows.

\section{Scalable Computational Framework for Intelligent EV Infrastructure Optimisation} \label{sec:method}

\subsection{Challenges}
Despite reformulation into a single-level structure, the PA problem remains computationally challenging due to two main factors: \emph{high dimensionality} of traffic distribution variables and \emph{mixed-integer} decision variables. Specifically, the driver distribution profile \( \mathbf{Q} \) scales with the complexity of the TN, requiring route-level and extended path-level decisions across all O-D pairs. Moreover, placement variables \( x_i \) are integer-valued, further complicating direct optimisation. As network size increases, solving the PA problem with standard Mixed-Integer Nonlinear Programming solvers becomes computationally demanding.

\subsection{Proposed Solution Approach: JPPO-DE Framework}\label{Sec: Approximation Approach}
To address these challenges, we propose the \textbf{Joint Placement and Pricing Optimisation under Driver Equilibrium (JPPO-DE)}, a two-level framework that combines problem decomposition with integer relaxation. JPPO-DE improves scalability and adaptability for large-scale EV infrastructure planning. The following subsections detail each level and its role in efficiently solving the PA problem.

\subsubsection{Level 1: Decomposition of Driver Behaviour Analysis}

The first approximation level leverages the current low EV driver rate, $\alpha= \frac{\sum_\omega\gamma_\omega}{\sum_\omega(\gamma_\omega+\gamma^0_\omega)}$, enabling a decomposition of driver behaviour into sequential stages rather than solving a fully coupled system. This significantly reduces the number of decision variables within each problem, improving computational efficiency and scalability.

The first level of JPPO-DE exploits the currently low proportion \( \alpha \) of en-route EV users during peak periods, enabling a sequential rather than fully coupled analysis. We first solve the \emph{PA-NCD} problem, where the PA predicts the network-wide equilibrium traffic distribution \( \mathbf{q}^{0}_\omega \) for NCDs, by excluding EV drivers and CSs. This step captures the steady-state background traffic conditions that will subsequently inform EV infrastructure planning.

Using the fixed background NCD flows obtained from PA-NCD, we then solve the \emph{PA-EV} problem. In this stage, the PA jointly optimises CS placement \( x \), pricing decisions \( y \), and EV driver distributions \( \mathbf{q}_\omega \), treating NCD traffic as fixed input. This decomposition substantially reduces the dimensionality of each optimisation step, improving computational efficiency while preserving solution feasibility.

When \( \alpha \) becomes significant, EV-induced congestion can substantially affect NCD traffic patterns. In such cases, an \emph{iterative refinement} procedure is applied: alternating between updating \( \mathbf{q}_\omega^0 \) and \( \mathbf{q}_\omega \) until reaching a stable network equilibrium. The convergence analysis of this iterative scheme is an important direction for future research. The full decomposition procedure is summarised in Algorithm~\ref{Algorithm1}.

\vspace{0.1cm}
\begin{algorithm}[t]
\caption{JPPO-DE - Level 1: Decomposition of Driver Behaviour Analysis}
\label{Algorithm1}
\begin{algorithmic}[1]
\Require Penetration rate: \( \alpha \)

\If{\( \alpha \) is low}
    \State \textbf{Stage 1: Solve PA-NCD}
    \State Solve the PA problem excluding EV drivers and CSs
    \State Obtain NCD traffic distribution \( \mathbf{q}^{0}_\omega \)
    \State \textbf{Stage 2: Solve PA-EV}
    \State Fix \( \mathbf{q}^{0}_\omega \) as background traffic
    \State Solve the PA- EV drivers to obtain \( x^* \), \( y^* \), \( \mathbf{q}_\omega \)
\Else
    \State \textbf{Iterative Refinement Procedure}
    \Repeat
        \State Fix \( \mathbf{q}^0_\omega \), solve PA-EV for \( x, y, \mathbf{q}_\omega \)
        \State Fix \( \mathbf{q}_\omega \), solve PA-NCD for \( \mathbf{q}^0_\omega \)
    \Until{Convergence to stable equilibrium}
\EndIf
\State \textbf{Return:} Optimal charger placement \( x^* \), pricing \( y^* \), and driver distributions \( \mathbf{Q} \)
\end{algorithmic}
\end{algorithm}
\begin{algorithm}[h]
\caption{JPPO-DE - Level 2: Integer Relaxation and Adjustment}
\label{algorithm 2}
\begin{algorithmic}[1]
\State \textbf{Relax} placement variables: allow \( x_i \in R \geq 0 \) for all \( i \in N \)
\State \textbf{Solve} the relaxed PA-EV problem using an NLP solver to obtain \( x^* \)
\State \textbf{Floor} each \( x^*_i \) to obtain integer parts \( x^{*'}_i = \lfloor x^*_i \rfloor \)
\State \textbf{Compute} the difference \(\Delta = \sum_i x^*_i - \sum_i x^{*'}_i\)
\State \textbf{Increment} the \( \lfloor \Delta \rfloor \) largest fractional parts by one to match the integer part of the relaxed sum
\State \textbf{Fix} the adjusted placement variables \( x^{*'} \)
\State \textbf{Re-solve} the PA-EV problem to optimise pricing \( y \) and EV driver distributions \( \mathbf{q}_\omega \)
\end{algorithmic}
\end{algorithm}
\subsubsection{Level 2: Integer Relaxation and Adjustment}

The second level of JPPO-DE improves scalability by relaxing the integer placement variables \( x_i \) to continuous values \( x_i \geq 0 \), allowing the PA-EV problem to be efficiently solved using a standard Nonlinear Programming (NLP) solver. The relaxed solution \( x^* \) is then adjusted to ensure integer feasibility: fractional values are floored, and the largest remainders are incremented until the total number of chargers matches the integer part of the relaxed optimal sum.

After adjusting placements, the PA-EV problem is re-solved with fixed \( x \) to optimise charging prices \( y \) and EV driver distributions \( \mathbf{q}_\omega \). This two-step approximation balances computational efficiency with practical feasibility. The full procedure is summarised in Algorithm~\ref{algorithm 2}.

\section{Experiments and Evaluation} \label{sec:numerical_results}
We evaluate our framework on the benchmark Sioux Falls transportation network ~\cite{leblanc1975}, a benchmark applied in transport research~\cite{bagl, wang2019, zardini2023}. The model enables public authorities to identify optimal charging locations, determine charger counts, set dynamic prices, and anticipate traffic distribution under strategic driver behaviour. Results demonstrate its effectiveness in addressing key ITS objectives, including resilience, efficiency, and scalability through integrated placement and pricing optimisation.

\subsection{Experimental Setup}

\begin{figure}
    \centering
    \includegraphics[width=1\linewidth]{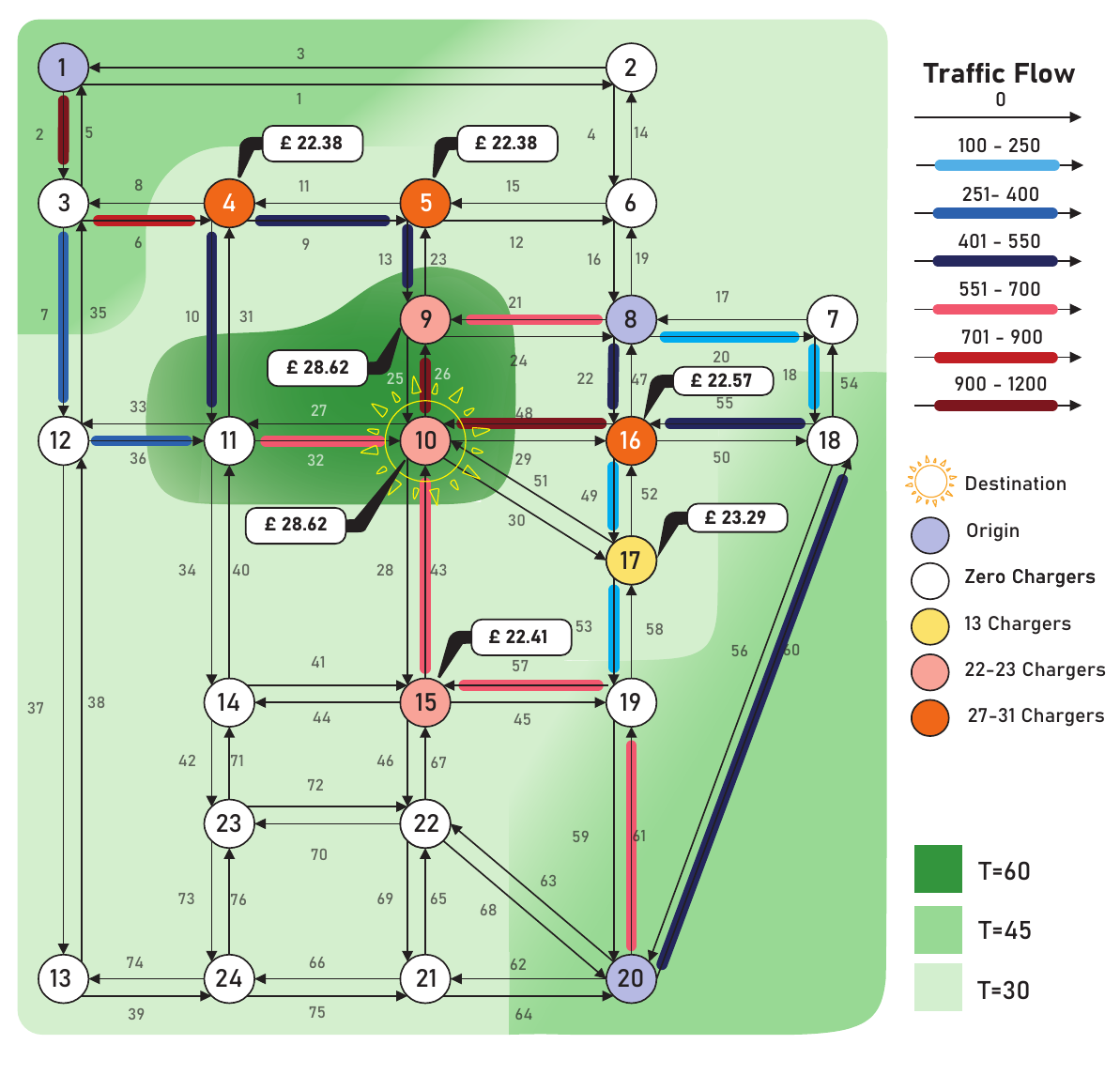}
    \caption{\textbf{Sioux Falls Network}: Solving PA model using JPPO-DE framework for peak-hours results in optimised EV Charging network- Reflecting traffic flow of drivers on links in equilibrium, charging fee and number of chargers in each node and different zones in terms of site rental \& maintenance costs $(T)$. The numbers on the links and nodes correspond to their respective identifiers.}
    \label{fig:Siouxfall}
\end{figure}

We evaluate our framework on the benchmark Sioux Falls transportation network (Fig.~\ref{fig:Siouxfall}), which includes 24 nodes and 76 links. Three representative O-D pairs (from residential zones to the city centre) simulate peak-hour demand. The scenario includes both NCDs (background traffic) and EV drivers who require en-route charging, with an EV share \(\alpha = 13\%\). Total traffic flow per O-D pair is 1150, resulting in 3450 drivers across the network. Experiments are implemented in Pyomo, using IPOPT for nonlinear programming. 


The model parameters are as follows: time-to-money conversion factor $\lambda = £25.12/\text{h}$~\cite{DfT2015}, uniform link capacity $c_l = 1000$, profit coefficient $\pi = 1.2$, and charger service rate $\mu = 4$. Site-specific maintenance costs $T_i$ are defined across three pricing zones, reflecting higher rental fees in the city centre. EVs are modelled as Nissan Leafs with a 24~kWh battery and 85\% charging efficiency~\cite{Channegowda2015}, requiring approximately 28.24~kWh per full charge. At a wholesale electricity rate of £0.27/kWh (uniform across sites), the energy cost per EV charge\footnote{The PA model optimises retail pricing while $e_i$ is treated as fixed input.} is set to $e_i = £7.5$. 

Note that while the current evaluation focuses on a single peak-hour scenario, the JPPO-DE framework can handle dynamic contexts with fluctuating O-D demand or network disruptions. By iteratively re-solving over time, it adapts to changes such as route shifts, traffic restrictions, CS outages, or fixed CS deployments, ensuring continued relevance in evolving ITS environments.

\subsection{Technique Application and Validation} \label{sec:technique}

\begin{figure}
    \centering
    \includegraphics[width=1\linewidth]{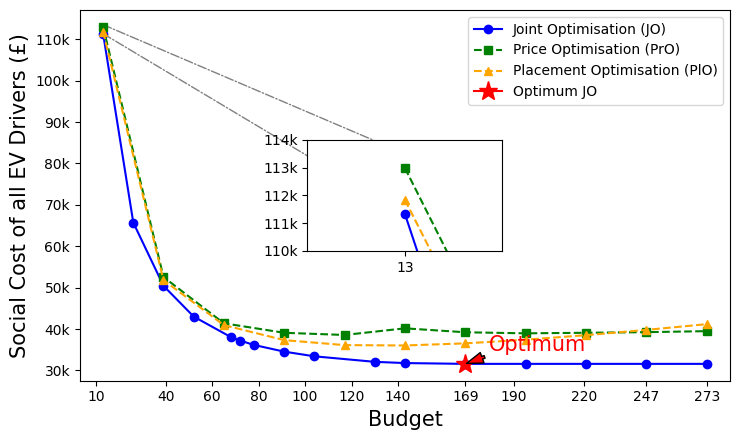}
    \caption{\textbf{Total Social Cost of EV drivers VS. Budget within 3 Optimisation Approaches:} JO consistently minimises total social cost across all budget levels. At the optimal budget of $B=169$, JO reduces costs by 16\% compared to PlO and 19\% compared to PrO, due to its ability to avoid over-installation in low-demand areas and overpricing in low-cost zones.}
    \label{fig:Sensitivity on B}
\end{figure}

We apply the JPPO-DE framework to jointly optimise charger placement, pricing, and driver flow distributions. Figure~\ref{fig:Siouxfall} illustrates the solution for the Sioux Falls network under a sufficiently large budget ($B \geq 169$), showing node-level charger quantities, prices, and total traffic flows (including both EV and NCD). This configuration achieves the minimum social cost while covering CS-related expenses and ensuring a 20\% profit margin. The results highlight that substantial charger installation is required even in high-cost zones ($T_i$), balanced by elevated pricing. As shown in Figure~\ref{fig:Sensitivity on B}, increasing the budget beyond $B = 169$ yields no further gains, as excessive installation in low-demand areas drives up costs without enhancing the user experience. The model ensures that charger numbers are capped at the optimal level, preventing over-installation.

\subsection{Comparison with State-of-the-Art Methods} \label{sec:comparison}

Following the approaches discussed in Section \ref{sec:literature}, we benchmark our proposed Joint Optimisation (JO) framework against two widely used simplified strategies: Pricing-only (PrO) and Placement-only (PlO). In PrO, chargers are evenly distributed and prices are optimised to manage demand, following the dynamic pricing approaches in prior work. In PlO, charger locations are optimised under a fixed uniform pricing scheme set to cover the highest operational cost with a marginal profit, as commonly done in siting-focused studies.

Unlike these sequential or single-focus methods, JO simultaneously optimises placement and pricing decisions, capturing their interdependence under realistic traffic conditions. As shown in Figure \ref{fig:Sensitivity on B}, JO consistently achieves lower total social cost across all budget levels. For instance, at the optimal budget threshold of $B=169$, JO reduces cost by 16\% relative to PlO and 19\% compared to PrO. This efficiency stems from JO’s ability to avoid over-installation in low-demand areas and prevent overpricing in low-cost zones. 

These results reinforce the practical value of JO for infrastructure planners, enabling more adaptive, demand-driven resource allocation. Robustness across varying budgets and traffic conditions is further demonstrated in the sensitivity analysis (Section \ref{sec:Sensitivity Analysis}).

\subsection{Approximation Accuracy}
We compared solutions from the relaxed continuous placement model with those from the integer adjustment (Algorithm \ref{algorithm 2}) to assess the impact of our two-level approximation. Across all scenarios, the maximum deviation in social cost was under 2\%, confirming that our approach delivers near-optimal results while maintaining computational efficiency.

\subsection{Sensitivity Analysis}\label{sec:Sensitivity Analysis}
\begin{figure}
    \centering
    \includegraphics[width=1\linewidth]{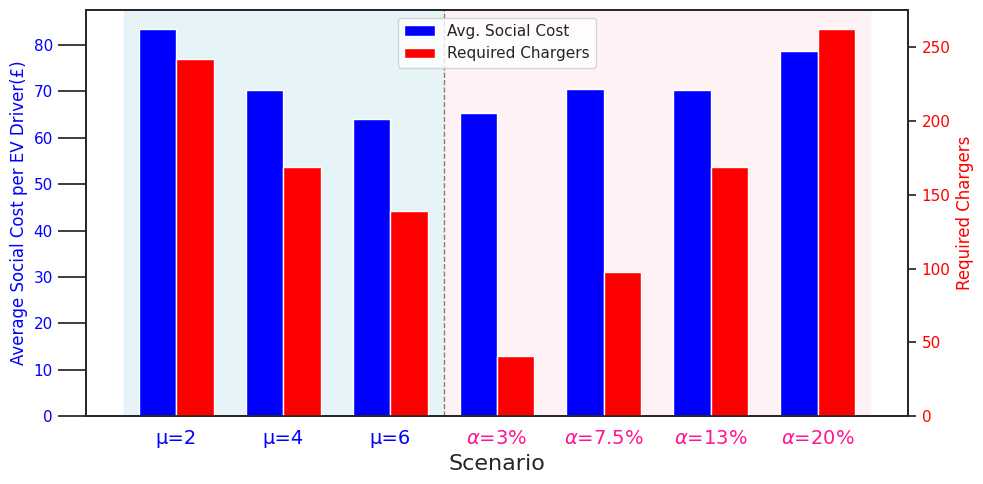}
    \caption{\textbf{Average Social Cost and Required Chargers vs. $\boldsymbol{\mu} \& \boldsymbol{\alpha}$}: \\ \textbf{Left}: As the charger service rate $\mu$ increases, assuming fixed $e$, queueing time drops, reducing the number of chargers needed and charging fees. \textbf{Right}: For each 1\% increase in $\alpha$, about 13 more chargers are required. Although the average social cost per EV rises gradually with greater penetration, expanding infrastructure alone may be inadequate. Additional measures, such as demand-side incentives, may be necessary to sustain system efficiency.}

    \label{alpha sensitivity}
\end{figure}

We conducted one-at-a-time sensitivity tests to assess robustness to key parameters. Beyond budget effects (Figure~\ref{fig:Sensitivity on B}), this subsection examines how charger service rate \(\mu\) and EV penetration \(\alpha\) impact infrastructure planning (Figure~\ref{alpha sensitivity}).

Charger speed \(\mu\): Increasing the service rate of chargers yields two main benefits: shorter queueing times for EV drivers and reduced infrastructure requirements. As shown in Figure \ref{alpha sensitivity} (left), faster chargers enable the same demand to be met with fewer stations, lowering both installation costs and user prices. Although this analysis assumes uniform energy demand, it clearly demonstrates the value of investing in high-capacity chargers to improve system efficiency.

EV penetration \(\alpha\): Figure \ref{alpha sensitivity} (right) shows that as en-route charging demand increases, the number of required stations rises by about 13 per 1\% increase in \(\alpha\). While the average social cost per EV increases slowly with higher penetration, infrastructure scaling alone may not suffice. Complementary policies, like demand-side incentives, may be needed to maintain system performance.

\subsection{Network Resilience Evaluation}

We evaluate the resilience of the JPPO-DE framework under CS disruptions, simulating failures where a subset of nodes lose power and their chargers become unavailable. In traditional fixed-design schemes, pricing decisions remain unchanged, and only driver flows adjust in response to the new network conditions. This mismatch between design and infrastructure leads to increased social costs.

\begin{figure}
    \centering
    \includegraphics[width=1\linewidth]{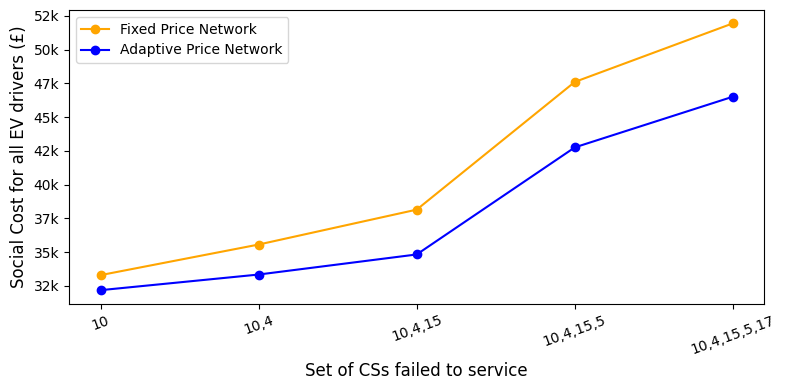}
    \caption{\textbf{Network Resilience Response to CS Failure:} Lower social costs emerged in the network when dynamic JPPO-DE was applied to adapt prices, compared to the fixed price model.}
    \label{network resilience}
\end{figure}

By contrast, JPPO-DE adapts to disruptions through dynamic price updates, re-optimising the charging tariffs to minimise system-wide costs given the modified CS availability. Figure~\ref{network resilience} compares the total social cost under both approaches for varying CS failure scenarios. Results show that JPPO-DE reduces social cost by up to 32\%  relative to the fixed-pricing scheme, highlighting its ability to improve system resilience through adaptive design.

\subsection{Scalability and Generalisability} 

The insights gained from our experiments support the extension of the JPPO-DE framework to larger transportation networks. Across various parameter settings, results consistently show that optimal solutions tend to concentrate charger installations at a limited number of strategically important nodes. This observation mitigates scalability concerns arising from the exponential growth in extended paths and optimisation variables with increasing network size.

To manage scalability, we adopt an iterative candidate-node approach: the model is first run on a limited subset of potential sites, and nodes with zero chargers are iteratively replaced by new candidates until all nodes are evaluated. This method significantly reduces the problem size at each iteration, enabling efficient optimisation for large-scale networks.
We validated this approach on multiple synthetic networks with up to 100 nodes, confirming that JPPO-DE yields feasible solutions within reasonable runtimes.

\section{Conclusion}

We presented an intelligent, congestion-aware optimisation framework for EV charging infrastructure planning that jointly addresses charger placement and pricing under realistic traffic dynamics. By modelling both road and station congestion through a coupled congestion game, our approach captures the strategic behaviour of EV and non-charging drivers within a unified transport network model.

Unlike prior studies that optimise siting or pricing in isolation, our framework enables public authorities to coordinate economic and spatial decisions simultaneously, supporting scalable, cost-effective deployment. The JPPO-DE algorithm offers a tractable solution to the resulting bi-level problem via behavioural decomposition and integer relaxation.

Experiments on a benchmark network demonstrate superior performance over single-parameter baselines, robust adaptation to changing budgets and EV penetration, and improved resilience to infrastructure disruption. The model generalises to larger networks through candidate-based iteration and offers clear guidance for policymakers and planners seeking to align mobility, efficiency, and sustainability goals.

Future work will extend the framework to support more realistic features, such as dynamic pricing, competitive ownership models, and heterogeneous vehicle behaviour and charging duration in real-world deployments.


\section{Acknowledgement}
This paper has been accepted for publication in the Proceedings of the IEEE 28th International Conference on Intelligent Transportation Systems (ITSC 2025). The final version will be available via IEEE Xplore.

\end{document}